\documentclass[pra,twocolumn]{revtex4}

\usepackage{amsmath}
\usepackage{graphicx}
\usepackage{pstricks}

\begin{document}

\title{Quasi phase matching for high order harmonic generation induced by the carrier-envelope phase.}

\author{Daniele Faccio$^{1,2}$, Carles Serrat$^{1,3}$, Jos\'e M. Cela$^4$, Albert Farr\'es$^4$,  Paolo Di Trapani$^{2,5}$,  Jens Biegert$^{1,6}$}

\address{$^{1}$ICFO-Institut de Ci{\`e}nces Fot{\`o}niques, Mediterranean Technology Park, 08860 Castelldefels, Barcelona, Spain\\
$^2$CNISM and Department of Physics and Mathematics, Universit\`a dell'Insubria, Via Valleggio 11, IT-22100 Como, Italy\\
$^3$ DTDI Universitat de Vic, Carrer de la Laura 13, 08500 Vic, Barcelona, Spain\\
$^4$ CASE Barcelona Supercomputing Center, Carrer Gran Capit\`a 2-4, 08034 Barcelona, Spain\\
$^5$ Virtual Institute for Nonlinear Optics, Centro di Cultura Scientifica Alessandro Volta, Via Olmo Via Simone Cantoni 1,  22100 Como, Italy \\
 $^{6}$ICREA Instituci\'o Catalana de Recerca i Estudis Avan\c cats, 08010 Barcelona, Spain
}
\email{ daniele.faccio@uninsubria.it}

\begin{abstract}
We report a novel quasi-phase matching technique for high-order harmonic generation in low-density gases. Numerical simulations show that in few-optical cycle pulsed Bessel beams it is possible to  control the pulse envelope and phase velocities which in turn allows to control the carrier-envelope phase  during propagation. The resulting oscillations in the peak intensity allow to phase-match the high-order harmonic generation process with a nearly two decade enhancement in the XUV power spectrum.  
\end{abstract}

\maketitle
Recent developments in ultrashort laser pulse technology have proved the viability and effectiveness of producing Extreme Ultra-Violet (XUV) radiation by High Order Harmonic Generation (HHG)  in low pressure gases. This process may be qualitatively understood through the three-step model \cite{kulander,corkum} by which the atomic potential is strongly modified by the impinging laser radiation such that the electron may tunnel out of the potential well. Under the external influence of the strong driving laser field the electron wave-packet will be first driven away and then pulled back to the atom. The recombination process will liberate the excess energy in the form of high energy XUV photons. Both the phase of the driving laser field and phase accumulated during propagation will finally determine how the XUV field coherently builds up so that in general, in order to achieve efficient HHG along a defined propagation direction phase matching requirements must be fulfilled \cite{balcou,pfeifer,KMIEEE}.
The propagation distance over which this occurs is defined as the coherence length $\ell_c=\pi/|\Delta K|$, where the total phase mismatch $\Delta K$ for HHG from the fundamental laser frequency $\omega$ to the $q^{\textrm{th}}$ harmonic may be written as
\begin{equation}\label{deltaK}
\Delta K =K_\textrm{disp} +K_\textrm{plasma}+K_\textrm{atomic}+K_\textrm{geometric}.
\end{equation}
The contribution due to gas dispersion is $K_\textrm{disp}=qk_{\textrm{laser}}-k_{\textrm{q}}$, where $k_{\textrm{laser}}$ and $k_{\textrm{q}}$ are the laser and Harmonic pulse wave-vectors, respectively. This term, which typically has positive sign, for low gas pressures (e.g. {$<$} 50 mbar) can be neglected  \cite{priori}. The plasma dispersion term may be written as $K_\textrm{plasma}=\omega_p^2(1-q^2)/2qc\omega$, where $\omega_p$ is the plasma frequency, and gives a negative contribution \cite{pfeifer}, while the atomic dipole phase term $K_\textrm{atomic}$,  related to the quantum paths of the electrons involved in the HHG process, in general does not have a fixed sign during propagation since it is proportional to the intensity gradient along the propagation direction, $dI/dz$ \cite{balcou,lewenstein}.
Finally, the geometric term $K_\textrm{geometric}$ depends on the shape of the driving laser pulse. For example the Gouy phase shift in a tightly focused Gaussian pump pulse will give a positive contribution $K_\textrm{geometric}=2(q-1)/b$ where $b$ is the confocal parameter whilst a pump pulse confined in the lowest order hollow-core waveguide mode will give a negative contribution $K_\textrm{geometric}=-u_{11}^2c/2\omega a^2$, where $u_{11}$ is the first zero of the $J_0$ Bessel function and $a$ is the fiber radius \cite{KMIEEE}.\\
 \indent Truncated Bessel beams, obtained by refocusing with a lens the output of a hollow core waveguide, have  been proposed and the specific phase profile, related to the geometry of the driving pulse, has shown to lead to improved HHG brightness \cite{nisoli_bessel}. More recently Pulsed Bessel Beams (PBB) without successive refocusing, have also been proposed to optimize phase-matching in the XUV region \cite{ale,auguste}. PBBs have the pulse wave vectors aligned along a conical surface such that the resulting interference pattern is characterized by a central intense peak with surrounding low-intensity rings \cite{pbb_def}. By changing  the angle $\gamma$ of the wave-vector cone it is possible to continuously tune the negative geometric term,  $K_\textrm{geometric}=-q(\omega/c)(1-\cos\gamma)$ \cite{ale} and thus achieve perfect phase-matching ($\Delta K\simeq 0$) for a wide range of operating parameters. Due to the negative sign of this last contribution, phase-matching is possible only in the presence of a positive material dispersion (e.g. at high gas pressures) that is larger than the negative plasma contribution.\\
\indent In the absence of the possibility to achieve perfect phase-matching the next best thing is so-called Quasi-Phase-Matching (QPM) by which the HHG process is periodically strongly reduced or eliminated at propagation distances corresponding to odd multiples of the coherence length $\ell_c$. This fruitful technique has been implemented by a number of means some of which involve a static modulation of the gas medium while others are based on a more flexible and tunable counter-propagating beam configuration \cite{KMIEEE,qpm1,qpm2,qpm3,qpm4,qpm5} . \\
\indent In this work we propose a novel technique to achieve QPM with few-cycle pulses by controlling the Carrier-Envelope-Phase (CEP) along propagation. This technique is very general and may in principle be applied to any nonlinear intensity-dependent process although here we consider the specific case of HHG. The underlying idea is very simple: when the pulse contains only a few optical cycles the instantaneous peak intensity $I_{\textrm{inst}}$ will become very sensitive to the CEP value, i.e. to the relative phase between the carrier-wave and the pulse envelope. For CEP=0 ($\pi/2$), $I_{\textrm{inst}}$ will be maximum (minimum). Since HHG is extremely sensitive to variations in the pulse intensity and phase, a periodic modulation of the instantaneous intensity and phase may lead to a quasi-phase-matched conversion process. \\
\begin{figure}[t]
\includegraphics[angle=0,width=8cm]{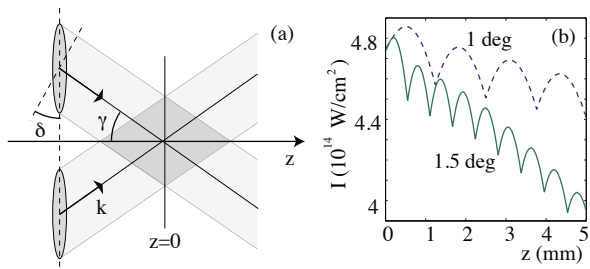} 
	\caption{ \label{fig:fig1} (in color online) (a) Scheme of pulsed Bessel beam geometry. $\gamma$ and $\delta$ are the Bessel con angle and the pulse front tilt angle, respectively. `z=0' indicates the position of the input condition for the numerical simulations. (b) shows the calculated  peak intensity evolution  on axis for the PBB for two different cone angles, $\gamma=1$, $1.5$ deg.}
\end{figure}
\indent This technique requires control over both the phase velocity, which determines how the carrier-wave propagates, and the group velocity, which determines how the pulse envelope propagates. Our proposal is to achieve this by use of PBBs. We first note that the PBB is just a particular case of a larger family of ``conical'' pulses for which the wave-vectors propagate along a cone with angle $\gamma$ and a pulse front that in general, may be tilted by an angle $\delta$ with respect to the wave-vector. The carrier-wave velocity, $v_{\phi}$, and the envelope peak velocity, $v_{p}$, along the propagation direction are related to the cone and tilt angles by \cite{valiulis}
\begin{eqnarray}
v_{\phi}&=&\omega_L/(K_L \cos\gamma) \label{vphi}\\
v_{p}&=&\cos\delta/K^{\prime}\cos(\gamma+\delta)\label{vp}
\end{eqnarray}
where it is implied that the $\omega_L$ and $K_L$ are the laser pulse carrier frequency and wave-vector, respectively and  $K^{\prime}=dK/d\omega$ is related only to material dispersion and is evaluated at $\omega_L$. In general we may thus independently choose the angles $\gamma$ and $\delta$ so that the phase and peak velocities are effectively two independent and tunable quantities. We focus our attention on the specific case $\delta=-\gamma$, i.e. of the PBB, as this allows sufficient control over the phase and peak velocity difference whilst remaining experimentally simple to access. For the case of a PBB, Eq.~(\ref{vp}) becomes $v_{p}=\cos\gamma/K^{\prime}$.  The difference between the phase and peak velocities will induce a beating in the pulse peak intensity with a periodicity that is twice that of the CEP and is a function of the cone angle $\gamma$, 
\begin{equation}\label{beating}
\Lambda(\gamma)=\cfrac{\pi}{\omega_L|v_\phi(\gamma)^{-1}-v_p(\gamma)^{-1}|}.
\end{equation}
So, for example, for a Bessel cone angle $\gamma=1$ deg, Eq.~(\ref{beating}) predicts a periodicity in the pulse intensity modulation  $\Lambda=1.28$ mm.\\
\indent In Fig.~\ref{fig:fig1}(a) we show a cross-section of the geometry of the PBB.  The tilted wavefronts propagate at an angle with respect to the propagation direction, $z$. The region in which they overlap is called the Bessel zone (darker shaded area in the figure): the line corresponding to z=0 indicates the position at which the PBB is considered in the present study as the input condition. Such a pulse and input condition may be obtained for example by illuminating a circular diffraction grating with a collimated Gaussian beam \cite{dyson,sochacki} or by nonlinear frequency conversion processes in nonlinear optical crystals \cite{opa_pbb}.\\
\begin{figure}[t]
\includegraphics[angle=0,width=8cm]{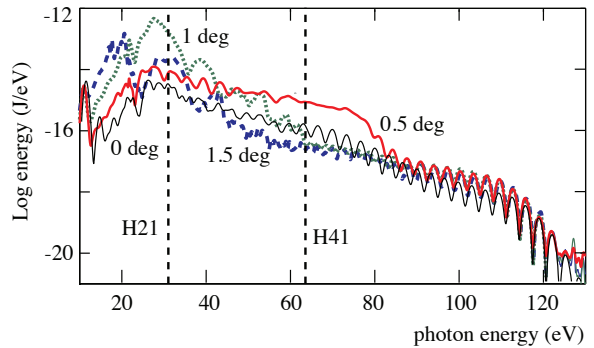} 
	\caption{ \label{fig:fig2} (in color online) XUV  relative spectral energy (in logarithmic scale) versus photon energy for three different PBB cone angles: dashed curve - 1.5 deg, dotted curve - 1 deg, solid curve - 0.5 deg. The thin black line indicated with 0 deg is the reference spectrum obtained with a Gaussian pump pulse that has a FWHM of 40 $\mu$m.
	 The spectral energy for 1 and 1.5 deg has been normalized to that of 0.5 deg at the cutoff (116.2 eV).
	The vertical dashed lines indicate the positions of the $21^{\textrm{st}}$ and $41^{\textrm{st}}$ harmonics, analyzed in more detail in Fig.~\ref{fig:fig3}.}
\end{figure}
\indent In order to verify the validity of this novel QPM technique we performed numerical simulations based on  a three dimensional propagation model in cylindrical coordinates 
using the nonadiabatic Strong-Field-Approximation \cite{SFA} to calculate the atomic response, as outlined in \cite{priori}. The computational code is fully parallelized with typical runs of 24 hours using 256 CPU in a high performance interprocess communication system.
The laser pulse central wavelength was chosen to be 800 nm, and the input field was written as  {$E=E_0J_0(r\omega_L\sin\gamma/c)\exp(-(r/\rho)^2)\exp(-(t/\tau)^2)\cos(\omega_L t)$} with a pulse duration 
at full-width-half-maximum (FWHM) of 5 fs (approximately 2 optical cycles). The Gaussian apodization in the transverse dimension was chosen to have a FWHM of 1.2 mm and the peak input intensity in 
all cases was $5\times10^{14}$ W/cm$^2$. HHG was performed in Neon gas at a pressure of 20 mbar over a propagation distance of 5 mm. These values were chosen so as to have very weak 
reabsorption and thus observe the effect in the simplest case. 
\begin{figure}[t] 
\includegraphics[angle=0,width=8cm]{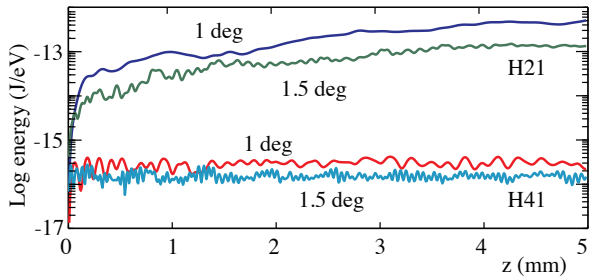}
	\caption{ \label{fig:fig3} (in color online) $21^{\textrm{st}}$ and $41^{\textrm{st}}$ harmonic relative spectral energy (in logarithmic scale) versus propagation distance for two different PBB cone angles, $\gamma=1$, $1.5$ deg.}
\end{figure}
 In Fig.~\ref{fig:fig1}(b) we show the evolution along the propagation direction of the peak intensity of the PBB for two different cone angles. It is the CEP-induced oscillations visible in the figure that drive the QPM process with periodicities that correspond to Eq.~(\ref{beating}). The slow decrease in the average intensity is due mainly to effective negative second order dispersion which arises due to the pulse front tilt in the PBB \cite{valiulis}. This negative dispersion at the pump pulse frequency may be partially compensated for by the positive material dispersion or eliminated by resorting to the use of more complicated non-dispersive pulses known as X-waves \cite{recami,valiulis}. However, not withstanding this slow intensity decrease, clear enhancement of phase-matching is observed as shown in Fig.~\ref{fig:fig2}. \\
 \indent Three different cases are considered, $\gamma = 0.5$, 1 and 1.5 deg. 
   The radially  integrated spectral energy of the XUV radiation (in Joules) per photon energy unit (in eV) is shown in Fig.~\ref{fig:fig2} for $\gamma = 0.5, 1, 1.5$ deg (the energies have been normalized to that of $\gamma = 0.5$ deg at the cutoff (116.2 eV) for clarity).
 All three PBBs show a nearly 2 decade enhancement of the XUV intensity with respect to  that of a Gaussian pulse. 
 Most interestingly it is also clear how changing the PBB cone angle gradually shifts the  phase-matched spectral region (with increasing photon energies for decreasing cone angles). Furthermore, 
 the bandwidth of the quasi-phase-matched spectral region increases with decreasing cone angle.\\
\begin{figure}[t] 
\includegraphics[angle=0,width=8cm]{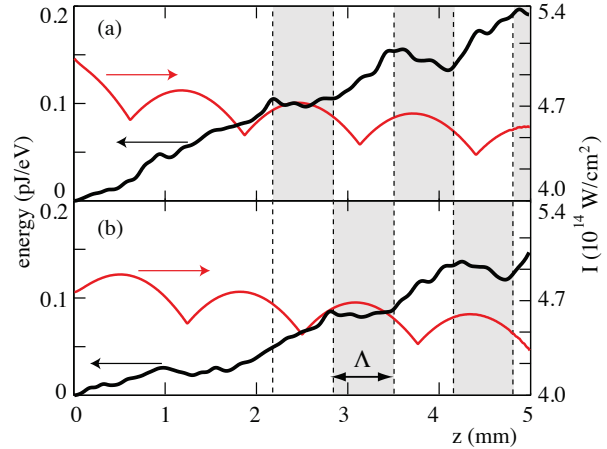}
	\caption{ \label{fig:fig4} 21$^{\rm st}$ (in color online) Black lines - harmonic energy for two different values of the initial CEP, 0 (a) and $\pi/2$ (b). The red lines show the laser pulse intensity (right axis) for the two cases. The shaded areas indicate the periodic plateau regions in the XUV QPM process. }
\end{figure}
 \indent In Fig.~\ref{fig:fig3} we show the XUV  relative spectral energy evolution versus the propagation distance for two different harmonics, the $21^{\textrm{st}}$ and the $41^{\textrm{st}}$ (whose spectral location is indicated by the dashed vertical lines in Fig.~\ref{fig:fig2}) and for two different PBB cone angles, 1 and 1.5 deg. These harmonics are chosen so that one is well within the phase-matched spectral region and the other is outside. As can be seen, the quasi-phase-matched $21^{\textrm{st}}$ harmonic intensity is clearly increasing over the whole propagation distance whilst the non quasi-phase-matched $41^{\textrm{st}}$ harmonic intensity continues to oscillate around a nearly constant value. This gives clear evidence of a phase-matching mechanism in the intensity-enhanced spectral regions.\\ 
\indent In Fig.~\ref{fig:fig4} we examine the XUV signal propagation in more depth. In particular we study the dependence of the $21^{\rm st}$ harmonic on the CEP oscillations and on the input CEP value for a PBB with $\gamma=1$ deg. In Fig.~\ref{fig:fig4}(a) the red line shows the laser pulse on-axis intensity for an initial CEP=0  and the black line shows the corresponding XUV energy evolution in linear scale. After an initial relatively monotonic growth, the XUV signal shows a periodic pattern in which the energy increases, reaches a plateau, increases once more, etc., hence clearly demonstrating the signature of QPM.  The shaded areas highlight the step-like behavior in the XUV signal that occurs with the same periodicity, $\Lambda\sim1.28$ mm, of the CEP induced oscillations and correspond to well-defined recurring positions in the periodic intensity pattern.
 In order to further investigate the influence of the CEP on phase-matching we performed an identical simulation with a different input CEP at z=0, CEP=$\pi/2$. This is shown in Fig.~\ref{fig:fig4}(b). The  resulting shift by half a period in the laser intensity modulation delays the onset of the XUV energy growth by  a corresponding half period. The following XUV signal evolution is then nearly identical to that observed for CEP=0 albeit with the same half-period shift to longer z. The harmonic generation is therefore strongly determined by the CEP induced oscillations and is locked not only to its periodicity but also to the precise value at each z. Finally, we note that the intervals in which the XUV energy increases do not coincide with the laser-pulse intensity maxima. This indicates that the periodic increase in the XUV signal cannot be simply ascribed to an increase in the laser intensity but is rather related to a periodic modulation in the phase contributions to the nonlinear process. It is precisely this periodic phase modulation that gives rise to the observed QPM in the harmonic energy.\\
\indent In conclusion we report a novel QPM mechanism by which HHG may be enhanced in a tunable XUV spectral region. This mechanism relies on the fact that few-cycle laser pulses with sufficiently different phase and group velocities will undergo periodic intensity oscillations due to CEP evolution during propagation. In the present work the phase and group velocities (and thus the CEP) are controlled by shaping the laser pulse into a pulsed Bessel beam and by then varying the Bessel cone angle. The resulting intensity oscillations are shown to give rise to QPM and a nearly two decade enhancement in the XUV intensity. This technique is very general and may be applied to other nonlinear frequency conversion regimes once a suitable method for controlling the CEP is adopted.
 
\indent The authors wish to acknowledge  support from the Consorzio Nazionale Inter-unversitario per le Scienze della Materia (CNISM), progetto INNESCO.  PDT acknowledges support from Marie Curie Chair project STELLA, Contract No. MEXC-CT-2005-025710. D.F. acknowledges support from Marie Curie grant, Contract No. PIEF-GA-2008-220085. JB and CS acknowledge  support from the Spanish Ministry of Education and  Science through its Consolider Program Science (SAUUL - CSD 2007-00013) as well as  through ÒPlan NacionalÓ (FIS2008-06368-C02-01/02). Computer resources and assistance provided by the Barcelona Supercomputing Center is acknowledged.\\


\begin{thebibliography}{99}
\bibitem{pbb_def}
A precise definition of the PBB requires that the cone angle, $\gamma$, varies with frequency such that the transverse wave-vector, $k_\perp=(\omega/c)\sin\gamma(\omega)$, is constant. In the text $\gamma$ is to be understood as the cone angle at the carrier frequency of the pulse.
\end{thebibliography}
\end{document}